\documentclass[%
 aps,
 jmp,%
 amsmath,amssymb,
 reprint,%
]{revtex4-1}

\usepackage{graphicx}
\graphicspath{{figures/}}
\usepackage{dcolumn}
\usepackage{bm}
\usepackage{subfigure}

\usepackage{soul}
\usepackage{amsmath,amssymb,amsfonts}
\usepackage{graphics}
\usepackage{color}
\usepackage{xcolor}
\definecolor{b}{rgb}{0,0,0}
\definecolor{rev}{rgb}{0,0,0}
\definecolor{rev2}{rgb}{0,0,0}

\usepackage{algorithm}
\usepackage{algorithmic}
\usepackage{enumitem}
\setlist[itemize]{leftmargin=*}

\usepackage{comment}
\usepackage{hyperref}
\usepackage{lipsum}
\usepackage{tabularx}
\usepackage{mathrsfs}
\usepackage{stackengine}

\begin{document}








\title{Interface learning of multiphysics and multiscale systems}



\author{Shady E. Ahmed}

\author{Omer San}%
 \email{osan@okstate.edu}
 
\author{Kursat Kara}
\affiliation{ 
School of Mechanical \& Aerospace Engineering, Oklahoma State University, Stillwater, OK 74078, USA.
}%

\author{Rami Younis}
\affiliation{ 
The McDougall School of Petroleum Engineering, The University of Tulsa, Tulsa, OK 74104, USA.
}%

\author{Adil Rasheed}%
\affiliation{ 
Department of Engineering Cybernetics, Norwegian University of Science and Technology, N-7465, Trondheim, Norway.
}%

\date{\today}

\begin{abstract}

Complex natural or engineered systems comprise multiple characteristic scales, multiple spatiotemporal domains, and even multiple physical closure laws. To address such challenges, we introduce an interface learning paradigm and put forth a data-driven closure approach based on memory embedding to provide physically correct boundary conditions at the interface. \textcolor{rev}{To enable the interface learning for hyperbolic systems by considering the domain of influence and wave structures into account, we put forth the concept of \emph{upwind learning} towards a physics-informed domain decomposition. The promise of the proposed approach is shown for a set of canonical illustrative problems.} We highlight that high-performance computing environments can benefit from this methodology to reduce communication costs among processing units in emerging machine learning ready heterogeneous platforms toward exascale era. 

\end{abstract}


\keywords{Interface Boundary, Digital Twin, Machine Learning, Hybrid Analysis and Modeling, Multiscale Systems} 
\maketitle


\begin{figure*}[ht]
\centering
\includegraphics[trim= 0 0 0 0, clip, width=0.95\textwidth]{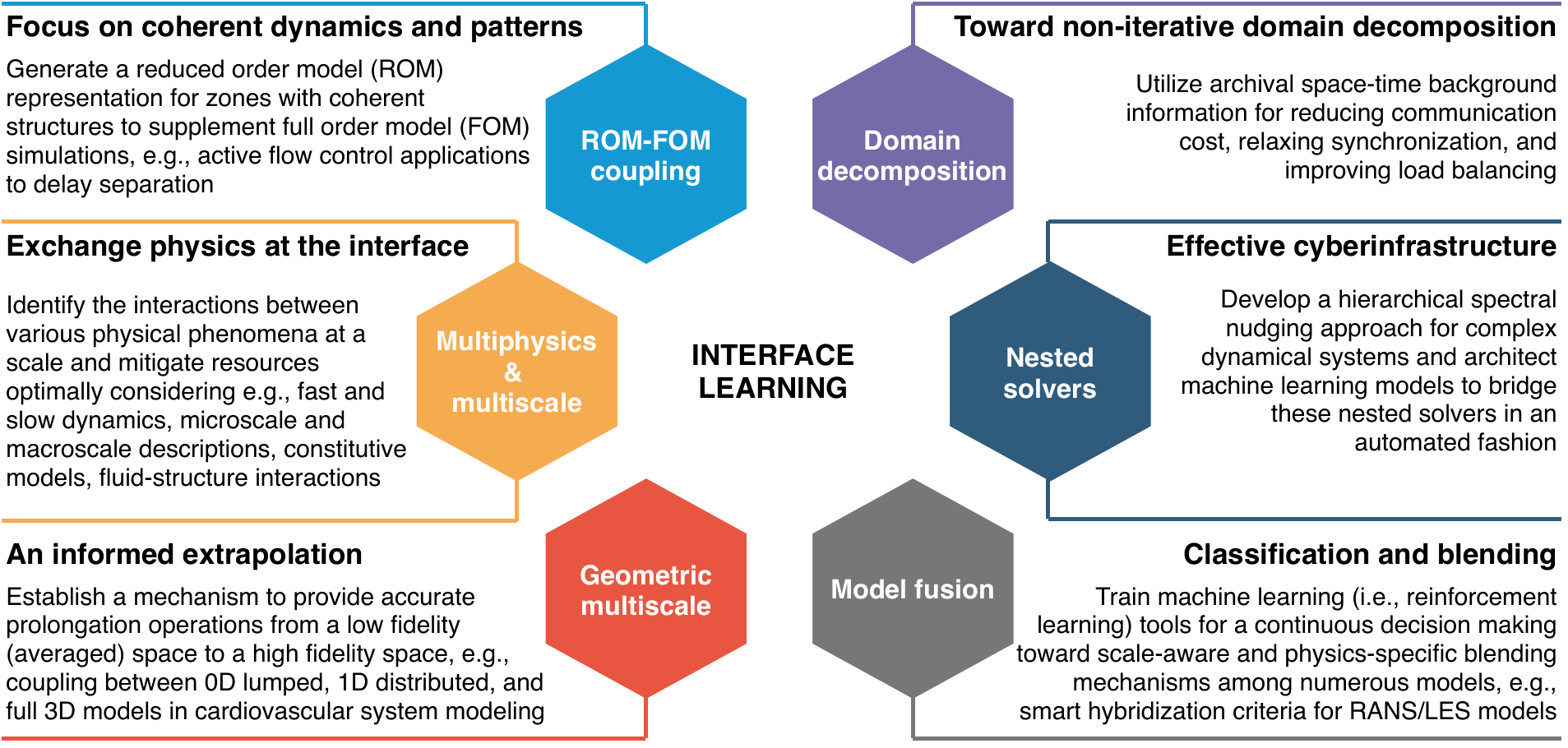}
\caption{Overview of the interface learning paradigm considering numerous scientific and engineering interpretations.}
\label{fig:lim}
\end{figure*}

\textcolor{rev}{Specification of boundary conditions is essential for the accurate solutions of mathematical models representing physical system \cite{roberts1992boundary}.} Moreover, numerical simulations of multiscale, multicomponent, multiphysics and multidisciplinary systems require additional treatment of interface boundary conditions among solvers or heterogeneous computational entities. Otherwise, in a naive implementation, the stiffest part of the domain dictates the overall spatial mesh resolution and time stepping requirements, making such simulations computationally daunting. Most of such hierarchical problems that incorporate some sort of information exchange can be put into the following six categories, explained with examples as follows:\\
\emph{1.~Reduced order model - full order model coupling:} With the emergence of digital twin like technologies, there is a demand for lighter models that can run in real time \cite{peherstorfer2018survey,naets2019multi}. In the context of weather prediction, the full order model (FOM) has been in use for a long time; however, they are incapable of modeling phenomena associated with scales smaller than what the coarse mesh can handle (like buildings and small terrain variations). These fine scale flow structures can be modeled using a much refined mesh but then the simulations become computationally intractable. To tackle this problem, a large variety of reduced order models (ROMs) are being developed. In order to make these ROMs realistic, there is a need to couple them to the FOM model so that the interface conditions (both in space and time) are exchanged between the FOM and ROM.\\
\emph{2.~Multiphysics and multiscale coupling:} Various flow dichotomies with a multiphysics coupling of interacting subsystems can be identified in many scientific and engineering applications \cite{keyes2013multiphysics}. For instance, in a gas turbine flow, the rotating parts and wall turbulence largely govern the flow within compressor and turbine sections. On the other hand, for the flow within the combustor, chemical reactions, heat release, acoustics, and the presence of fuel spray come into play. Thus, a simulation of the flow within the combustion chamber is significantly more expensive and demanding than other sections. It would require a finer and more sophisticated mesh, smaller time step, and less numerical simplifications. Therefore, using a unified global solver for the whole system would be either too expensive (matching the level of the fidelity required for the more complex part), or unacceptably inaccurate (following the level of fidelity required for the inexpensive part). Instead, multiple solvers are usually utilized to address different parts \cite{shankaran2001multi,xu2020reduced,chen2018boundary}, and information is transferred between solvers. Another common example in this category might be the use of a particle based approach in part of the domain, while using a continuum approach in the rest of the domain \cite{o1995molecular,pawar2020interface}.\\
\emph{3.~Geometric multiscale:} One example is the blood flow in the whole circulatory system which is mathematically described by means of heterogeneous problems featuring different degrees of detail and different geometric dimensions that interact together through appropriate interface coupling conditions. Proper exchange of interface conditions between models operating at different geometric approximations opens altogether new vistas for biofluids simulations \cite{quarteroni2003analysis,passerini20093d,quarteroni2016geometric}. Such multi-dimension modeling has been also promoted in porous media flows \cite{nordbotten2017modeling,boon2018robust,nordbotten2019unified}.  \\
\emph{4.~Model fusion:} Turbulence modeling generally requires an apriori selection of the most suited model to handle a particular kind of flow. However, it is seldom that one model is sufficient for different kind of zones in the computational domain. To alleviate this problem, hybrid and blending models have been extensively utilized to lift technical barriers in industrial applications, especially in settings where the Reynolds-averaged Navier-Stokes (RANS) approach is not sufficient and large eddy simulation (LES) is too expensive  \cite{sagaut2013multiscale,fadai2010seamless,shur2008hybrid}. The approach can be extended to blend any number of turbulence models provided the exchange of information at the interface can be accurately modeled \cite{maulik2019sub}.\\
\emph{5.~Nested solvers:} To decrease the computational cost required for an accurate representation of the numerous interconnected physical systems, e.g., oceanic and atmospheric flows, several classes of nested models have been developed and form the basis of highly successful applications and research at numerous weather and climate centers. Enforcing consistent flow conditions between successive nesting levels is also considered one form of interface matching. For example, a spectral nudging approach has been successfully implemented to force the large-scale atmospheric states from global climate models onto a regional climate model \cite{waldron1996sensitivity,von2000spectral,radu2008spectral,miguez2004spectral,rockel2008dynamical,schubert2017optimal}.\\
\emph{6.~Domain decomposition:} Since various zones in multiscale systems are connected through interfaces, data sharing, and communicating consistent interface boundary conditions among respective solvers are inevitable \cite{tang2020review}. 
Likewise, multirate and locally adaptive stepping methods can yield a mismatch at the space-time interface, and simple interpolation or extrapolation might lead to solution divergence or instabilities \cite{gander2013techniques}. 
An analogous situation usually occurs in parallel computing environments with domain decomposition and distribution over separate processors with message passing interface to communicate information between processors. The heterogeneity of different processing units creates an asynchronous computational environment, and the slowest processors will control the computational speed unless efficient load-balancing is performed \cite{donzis2014asynchronous,mittal2017proxy}. 

In short we can conclude that developing novel methodologies to model the information exchange at the interface will have far reaching impacts on a large variety of problems as shown in Fig.~\ref{fig:lim}. To this end, the current letter puts forth an approach based on memory embedding via machine learning to provide physically correct interfacial conditions. In particular, the proposed technique relies on the time history of local information to estimate consistent boundary conditions at the sub-domain boundaries without the need to resolve the neighboring regions (on the other side of the interface). It enables us to focus our computational resources on the region or scales of interest. We first present proof-of-concept computations on a bi-zonal one-dimensional Burgers' problem to showcase the proposed approach's promise for stiff multiscale systems. \textcolor{rev}{Moreover, we demonstrate that upwinding ideas can be easily incorporated in the interface learning framework to make the proposed approach physically more consistent with the underlying characteristics information and wave structures in hyperbolicity dominated systems. With this in mind, we then propose the \emph{upwind learning} approach toward establishing an eclectic framework for physics-informed data-driven domain decomposition approaches. The efficacy of the upwind learning is revealed using a set of canonical problems, including the hyperbolic Euler equations and pulsed flow equations.} We also highlight that high-performance computing environments can benefit from this methodology to reduce communication costs among processing units.

\emph{Two-component system} ---
As a \textcolor{rev}{first} demonstration of interface learning, we consider an application to the one-dimensional (1D) viscous Burgers problem. It combines the effects of viscous diffusion, friction, and nonlinear advection, and thus serves as a prototypical test bed for several numerical simulations studies. In order to mimic multiscale/multiphysics systems, we suppose the domain consists of two distinguishable zones corresponding to different physical parameters as follows,
\begin{align} \label{eq:brg}
    \dfrac{\partial u}{\partial t} &+ u \dfrac{\partial u}{\partial x} = \nu \dfrac{\partial^2 u}{\partial x^2} - \gamma u, \\
    (\nu, \gamma) &= \begin{cases}
            (10^{-2},0) \quad \text{for } 0\le x \le x_b\\
            (10^{-4},1) \quad \text{for } x_b < x \le 1,
            \end{cases}
\end{align}
where $x_b$ is the spatial location of the interface. In a naive implementation, a numerical solution of this problem would imply the use of a grid resolution and time step corresponding to the stiffest part (i.e., the left zone in this case) all over the domain unless we adopt an implicit scheme which is unconditionally stable but requires a nonlinear solver typically at each time step. Certainly, this puts an excessive and unnecessary computational burden. For instance, if we opt to using a spatial resolution of $4096$ grid spacings with a simple forward in time central in space (FTCS) finite difference scheme, the maximum time step that can be used in the left zone is approximately ~$2.5\times10^{-6}$ (i.e., $\delta t \leq \delta x^2/(2 \nu)$ based on von Neumann stability analysis). On the other hand, the right zone gives the flexibility of using two orders of magnitude larger time step. However, resolving the whole domain simultaneously would dictate the smaller time step, even if we are only interested in the right zone. A similar scenario would take place in multicomponent systems with varying spatial grid resolutions, where a unified resolution all over the domain becomes unpractical. Thus, we explore the introduction of a memory embedding architecture to enable resolving the zone of interest independently of the rest of the domain.

\begin{figure}
\centering
\includegraphics[trim= 0 0 0 0, clip, width=0.48\textwidth]{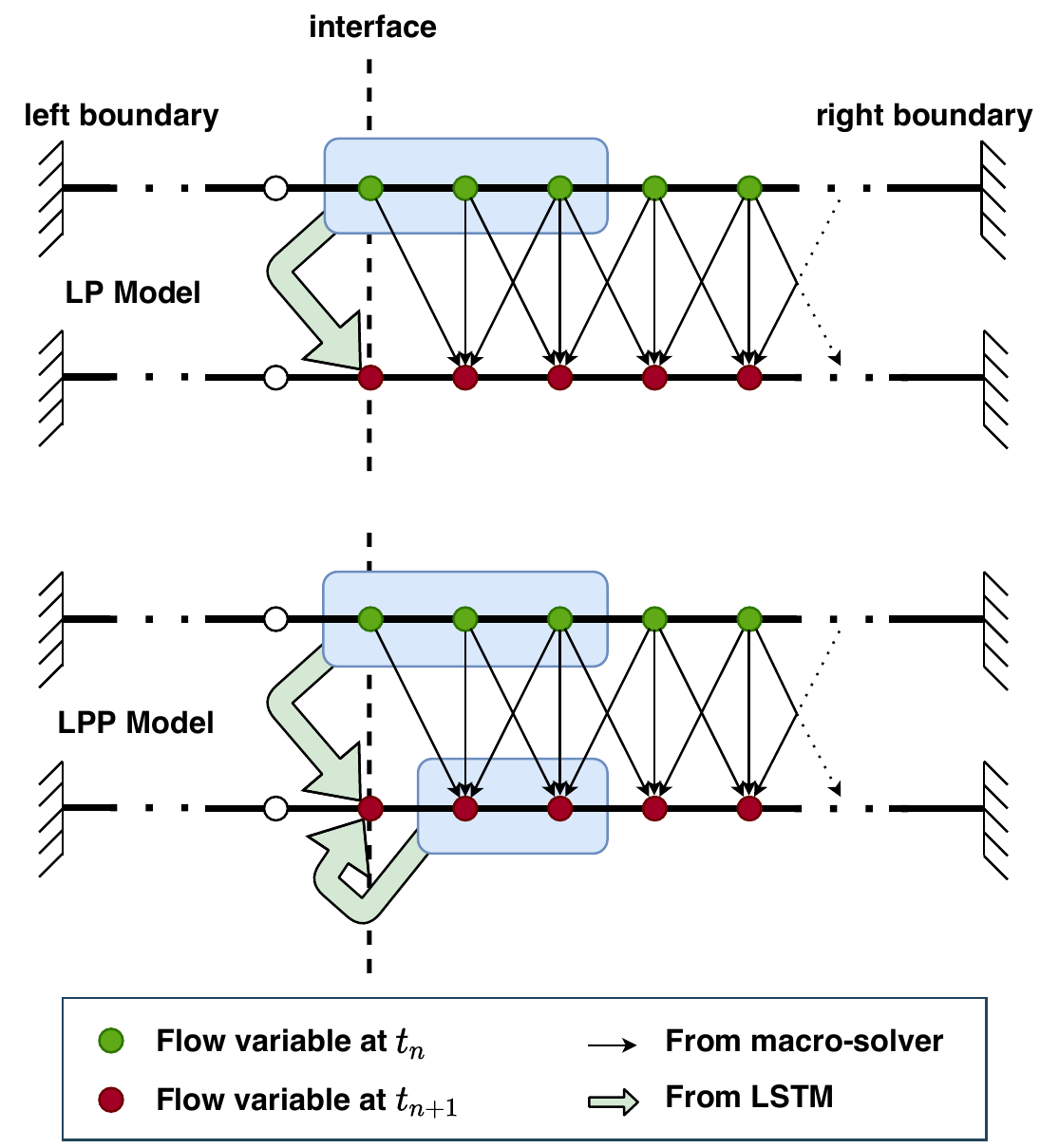}
\caption{Different models to utilize LSTM mapping for learning boundary conditions at interface}
\label{fig:models}
\end{figure}

\emph{Memory embedding of interface boundaries} --- 
For machine learning applicability, a pattern must exist and most fluid flows are dominated by coherent structures. Thus, our underlying hypothesis is the existence of a dynamical context or correlation between the time history of flow features at the interface in addition to the interactions with its one-sided neighbors (i.e., $u(x_b,t_n)$, $u(x_b+\delta x,t_n)$, $u(x_b+2\delta x,t_n)$, $\dots$), and the future state at the interface (i.e., $u(x_b,t_n+\delta t)$). This corresponds to the \emph{Learn from Past} (LP) model in Fig.~\ref{fig:models}. Since we incorporate a fully explicit time stepping scheme in our simulation, the interface neighboring points might be evolved in time before the interface condition is updated \textcolor{rev}{(e.g., using locally-frozen boundary conditions)}. Thus, a variant of the LP model based on a combination between old and updated values, namely the \emph{Learn from Past and Present} (LPP) model, can be utilized as well. Furthermore, we extend this mapping to take into account the time history dependence in a non-Markovian manner through the adoption of recurrent neural networks. Those exploit an internal state feature that reserves information from past input to learn the \emph{context} to improve and refine the output. For the neural network architecture, we use a simplistic long short-term memory (LSTM) of two layers, 20 neurons each. Although more sophisticated ML architectures and/or numerical schemes  might be utilized (e.g., \cite{rasheed2020digital,bowers2012improved,manica2011enabling,borggaard2006approximate}), the main objective of the present study is to emphasize the potential of neural networks to advance computational fluid dynamics (CFD) simulations for multiscale and multicomponent systems.

\emph{Proof-of-concept results} --- For the demonstration and assessment of the introduced methodology of interface learning, we \textcolor{rev}{first} consider two examples of varying complexity \textcolor{rev}{for the 1D Burgers problem with quadratic nonlinearity and Laplacian dissipation defined in Eq.~\ref{eq:brg}. We then introduce and discuss the need for \emph{upwind learning} to provide physically-aware domain decomposition to address hyperbolicity-dominated systems.} For interface learning, we consider two schemes/models for the training as illustrated in Fig.~\ref{fig:models} to learn the dynamics at the internal boundary separating the two compartments.

\begin{figure}[htbp!]
\centering
\includegraphics[trim= 0 0 0 0, clip, width=0.49\textwidth]{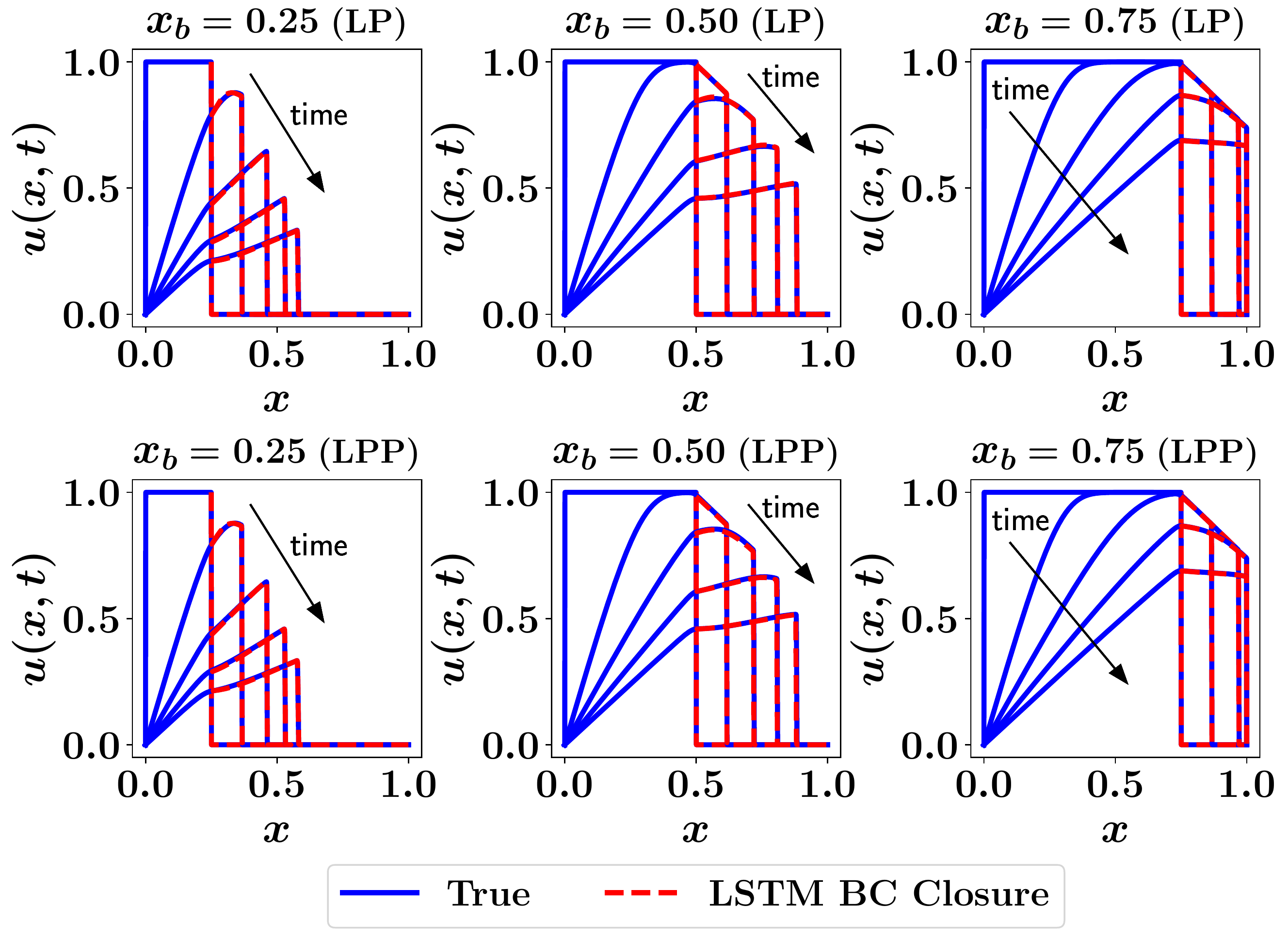}
\caption{Results for LSTM boundary condition closure for different values of $x_b$. Predicted velocity fields are shown at $t \in \{0.0, 0.25, 0.50, 0.75, 1.0\}$.}
\label{fig:res1}
\end{figure}

\emph{Example 1: travelling square wave}. In this first example, we address the problem of a travelling square wave, where the initial condition is defined with an amplitude of 1 in the left zone (i.e., $0\le x \le x_b$), and zero in the right zone as below,
\begin{align*}
    u(x, 0) &= \begin{cases}
            1 \quad \text{for } 0\le x \le x_b\\
            0 \quad \text{for } x_b < x \le 1.
            \end{cases}
\end{align*}
In other words, the interface is placed exactly at the discontinuity location of the initial propagating wave. So, the wave is guaranteed to instantaneously enter the right zone once the flow is triggered. We solve the presented viscous 1D Burgers problem for a time span of $[0,1]$ using a time step of $2.5\times 10^{-6}$ to resolve the whole domain $x\in [0,1]$ over a spatial grid resolution of $4096$. For external boundary conditions, we assume zero Dirichlet boundary conditions (i.e., $u(0,t)=u(1,t)=0$). Data snapshots are stored every 100 time steps (corresponding to the coarse time step of $2.5\times 10^{-4}$). In particular, we generate data for $x_b \in \{1/8, 2/8, 3/8, 4/8, 5/8, 6/8, 7/8 \}$, and we use field data at $x_b \in \{1/8, 3/8, 5/8, 7/8 \}$ for training and reserve the remaining cases for the out-of-sample testing.

For interface learning discussion, we consider the truncated 1D domain, where $x_b \le x \le 1$, and resolve the flow dynamics in this portion using a coarse time step of $2.5 \times 10^{-4}$, thus denoted the macro-solver here. We adopt the LSTM learning to update the left boundary condition (i.e., at $x=x_b$). For the right boundary (i.e., at $x=1$), we keep the standard zero Dirichlet conditions. To enhance the neural network performance, we augment the input vector with the spatial and temporal information as well. In other words, LP Model can be interpreted as the mapping $u(x_b,t_n+\delta t) = G_1\big( u(x_b,t_n), u(x_b+\delta x,t_n), u(x_b+2\delta x,t_n), \dots, x_b, x_b+\delta x, x_b+2\delta x, \dots, t_n \big)$, while LPP Model learns the map of $u(x_b,t_{n}+\delta t) = G_2\big(u(x_b,t_n), u(x_b+\delta x,t_n), u(x_b+2\delta x,t_n), \dots, u(x_b+\delta x,t_n+\delta t), u(x_b+2\delta x,t_n+\delta t), \dots, x_b, x_b+\delta x, x_b+2\delta x, \dots, t_n, t_n+\delta t \big)$. 

We compare the predicted velocity field within the truncated domain using the proposed LSTM boundary condition (BC) closure approach with respect to the true solution obtained by solving the whole domain. We note here that the LSTM BC closure results are based on utilizing the macro-solver (i.e., using a time step of $2.5\times10^{-4}$), while the true solution is obtained by adopting the micro-solver (i.e., using a time step of $2.5\times10^{-6}$). The spatio-temporal evolution of the velocity field for $x_b\in \{1/4, 2/4, 3/4\}$ is shown in Fig.~\ref{fig:res1}, where $x_b$ is the location of the interface. We note that Fig.~\ref{fig:res1} corresponds to a three-point stencil for the LSTM mapping. In other words, the LP model uses values of $u(x_b,t_n)$, $u(x_b+\delta x,t_n)$, and $u(x_b+2\delta x,t_n)$ for the prediction of $u(x_b,t_n+\delta t)$, while the LPP model uses  $u(x_b,t_n)$, $u(x_b+\delta x,t_n)$, $u(x_b+2\delta x,t_n)$, $u(x_b+\delta x,t_n+\delta t)$, and $u(x_b+2\delta x,t_n+\delta t)$. Visual results advocate the capability of the presented approach of predicting accurate values for the interface boundary condition at different times. For more quantitative assessment, we also compute the resulting root mean-squares error (RMSE) defined as
\begin{equation}
    \text{RMSE} = \sqrt{ \dfrac{1}{N_t N_x} \sum_{n=1}^{N_t} \sum_{i=1}^{N_x} \big( u^{\text{T}}(x_i,t_n) - u^{\text{P}}(x_i,t_n) \big)^2},
\end{equation}
where $u^{\text{T}}$ is the true velocity field, and $u^{\text{P}}$ represents the predictions by the LSTM BC closure approach. In the above formula, $N_x$ stands for the number of grid points involved only in the truncated domain. In other words, it considers only the flow field values within $[x_b,1]$.

The RMSE values of the LSTM BC closure predictions using a two- and three-point stencils are documented in Table~\ref{table:RMSE1} using the LP and LPP models. Quantitative results imply that the LP model is giving slightly better results than the LPP model. We believe that this behavior is because the LP model is more consistent with the adopted explicit numerical scheme, where the time evolution relies solely on the \emph{old} values of the flow field. Moreover, this might be attributed to the sub-optimal architecture we use for the LSTM. Although we found that results are not very sensitive to the given \emph{hyper-parameters}, further tuning might be required to provide \emph{optimal} performance. We also see from Table~\ref{table:RMSE1} that an increase in the stencil size from 2 to 3 improves results. Nonetheless, a 2-point stencil mapping still provides acceptable predictions, confirming the validity and robustness of the LSTM memory embedding skills to yield physically consistent and accurate state estimates at the interface using local information, and may hold immense potential for designing ML-ready predictive engines in physical sciences.

\begin{table}[htbp!]
\caption{RMSE of LSTM boundary condition closure results using different models with two-point and three-point mapping.}
\centering
\begin{tabular}{p{0.06\textwidth} p{0.1\textwidth} p{0.1\textwidth}  p{0.1\textwidth} p{0.1\textwidth} }  
\hline
  & \multicolumn{2}{c}{LP Model} &  \multicolumn{2}{c}{LPP Model} \\
$x_b$ &  2 Points & 3 Points & 2 Points & 3 Points \\
\hline \smallskip  \\
$0.125$   & $1.5\times10^{-2}$ & $2.9\times10^{-3}$ & $2.6\times10^{-2}$  & $1.4\times10^{-2}$  \\ 
$0.250$   & $4.4\times10^{-2}$ & $3.5\times10^{-3}$ & $2.8\times10^{-2}$ & $3.4\times10^{-3}$  \\ 
$0.375$   & $8.4\times10^{-3}$ & $3.6\times10^{-3}$ & $2.0\times10^{-2}$ & $1.6\times10^{-2}$  \\ 
$0.500$   & $2.9\times10^{-2}$ & $2.6\times10^{-3}$ & $1.7\times10^{-2}$ & $2.3\times10^{-2}$  \\ 
$0.625$   & $6.4\times10^{-3}$ & $3.7\times10^{-3}$ & $1.9\times10^{-2}$ & $1.3\times10^{-2}$  \\ 
$0.750$   & $1.0\times10^{-2}$ & $5.4\times10^{-3}$ & $1.2\times10^{-2}$ & $1.7\times10^{-3}$  \\ 
$0.875$   & $2.1\times10^{-3}$ &  $1.8\times10^{-3}$ & $6.7\times10^{-3}$ & $4.4\times10^{-3}$   \\
\hline
\end{tabular}
\label{table:RMSE1}
\end{table}

\begin{figure}[htbp!]
\centering
\includegraphics[trim= 0 0 0 0, clip, width=0.49\textwidth]{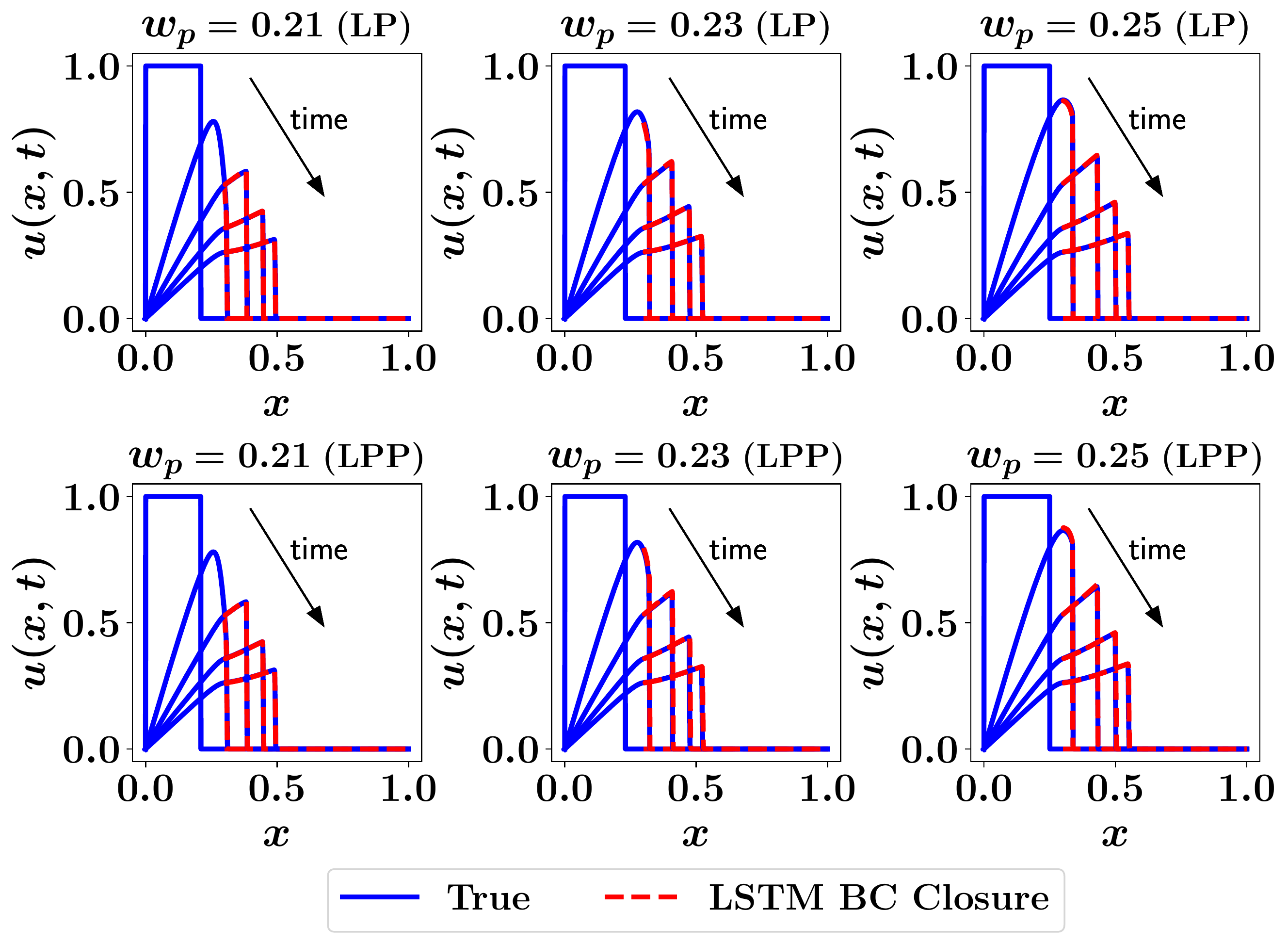}
\caption{Results for LSTM boundary condition closure for the pulse problem using different values of $w_p$. Predicted velocity fields are shown at $t \in \{0.0, 0.25, 0.50, 0.75, 1.0\}$.}
\label{fig:res2}
\end{figure}

\emph{Example 2: pulse problem}. In a second example of increasing complexity, we study the evolution of a pulse wave completely contained in a portion of the left region. The initiation of flow dynamics in the truncated domain is controlled by the interplay between advection, diffusion, and friction in different regions. Specifically, we consider an initial condition of a pulse, completely contained in the left region and study its propagation and travel from the left to right compartments. In particular, the initial pulse can be represented as
\begin{align*}
    u(x, 0) &= \begin{cases}
            1 \quad \text{for } 0\le x \le w_p\\
            0 \quad \text{for } w_p < x \le 1,
            \end{cases}
\end{align*}
where $w_p$ is the pulse width. For illustration, we store results corresponding to 7 varying pulse widths as $w_p \in \{0.20, 0.21, 0.22, 0.23, 0.24, 0.25, 0.26\}$. \textcolor{rev}{The same numerical schemes and resolutions of Example 1 are adopted here.} Data corresponding to $w_p \in \{0.20, 0.22, 0.24, 0.26\}$ are used for training and validation, while we assign $w_p \in \{0.21, 0.23, 0.25\}$ for out-of-sample testing. For interface, we consider a fixed interface location at $x_b=0.3$ (i.e., on the right of the largest pulse width). This is to let the interplay between the different interacting mechanisms (i.e., advection, diffusion, and friction) to come into effect \emph{before} the pulse travels into the truncated zone. Thus the state at the interface is more dependent on the flow dynamics in \emph{both} domain partitions. Since the pulse width is a key factor in this problem setting, we augment our input vector with $w_p$ as well. For this particular example, we found that enforcing higher memory embedding is crucial in providing accurate results. Specifically, we adopt a sliding window of a three-time step (also called a lookback of 3) in our LSTM implementation \cite{rahman2019nonintrusive}. Results for the LP and LPP schemes are shown in Fig.~\ref{fig:res2} using 3-point mapping. We find that both schemes can sufficiently learn the interface dynamics and accurately predict its condition at out-of-sample settings.

\textcolor{rev}{Although the wave in the previous examples moves from left to right, both LP and LPP are able to predict the interface conditions from the right-sided neighbors. In other words, the upstream conditions are inferred from the time history of the downstream flow. However, we highlight that this was greatly feasible due to the significant dissipative (viscous) effects that enable the rapid dissipation of information across the whole domain. Therefore, as soon as the solver is initialized within the right sub-domain, the incoming wave is already \emph{felt} downstream. That is why a deeper sliding window was needed in Example 2 to allow the effect of the pulse to pass to the right zone. On the other hand, in the hyperbolic limit, this cannot be established. For instance, if we look at the linear advection problem given as $u_t + a u_x = 0$ (where $a$ is the constant wave speed), we know that the solution can be written as $u(x,t)= u(x-at,0)$. This means that the specification of the interface condition $u(x_b,t)$ relies on the information from the direction where the wave is coming from (e.g., for positive $a$, we need information from the left sided neighbors).}

\textcolor{rev}{This is one limitation of the presented interface learning framework. In order to mitigate and treat this issue, we put forth an \emph{upwind learning} methodology to enforce physics into the learning process. For such, the domain decomposition procedure is performed to allow upwind learning from the upstream neighbors. In other words, instead of solving the whole domain, we replace the insignificant downstream subcomponents by an LSTM architecture and infer their effects from the time history of upstream flow dynamics. \textcolor{rev2}{Thus, the main purpose of the upwind learning framework is to account for the effects of downstream domain with an ML model that provides the interface condition in order to be able to solve for the upstream domain.} We illustrate the upwind learning methodology using a gas dynamics flow problem governed by the hyperbolic Euler equations as well as a pulsatile flow through a network of branched elastic tubes. }

\emph{Example 3: Euler equations of gas dynamics}. In this example, we apply the proposed upwind learning concept on the Sod's shock tube problem \cite{sod1978survey}, \textcolor{rev2}{considering  a long one-dimensional (1D) tube, closed at its ends with a thin diaphragm dividing the tube into two regions.} The governing one dimensional Euler equations can be written in a conservative form as below
\begin{align}
 \dfrac{\partial }{\partial t} \begin{bmatrix} \rho \\ \rho u \\ \rho e\end{bmatrix}   + \dfrac{\partial}{\partial x} \begin{bmatrix} \rho u \\ \rho u^2 + p \\ \rho u h \end{bmatrix}  = 0,
\end{align}
where $\rho$ is the density, $u$ is the horizontal component of the velocity, $e$ is the internal energy, $p=\rho(\gamma-1)(e-u^2/2)$ is the pressure and $h=e+p/\rho$ is the static enthalpy. Here, the ratio of specific heats is set to $\gamma=7/5$. \textcolor{rev2}{The two regions are initially filled with the same gas, but with different thermodynamic parameters, as follows,}
\begin{align*}
    (\rho, p, u )  &= \begin{cases}
            (1.0,1.0,0.0) \quad \text{for } 0\le x \le 0.5\\
            (0.125, 0.1, 0.0) \quad \text{for } 0.5 < x \le 1,
            \end{cases}
\end{align*}
with Dirichlet boundary conditions at $x=0$ and $x=1$. Data are collected for the time evolution from $t=0$ \textcolor{rev2}{to} $t=0.2$ using a time step of $10^{-4}$ and spatial step size of $2.5\times 10^{-3}$. \textcolor{rev2}{As a result of the diaphragm breakage, a contact discontinuity and a shock wave move from the left the right, while a rarefaction (expansion) wave moves from the right to the left.}

\begin{figure}[htbp!]
\centering
\includegraphics[trim= 0 0 0 0, clip, width=0.49\textwidth]{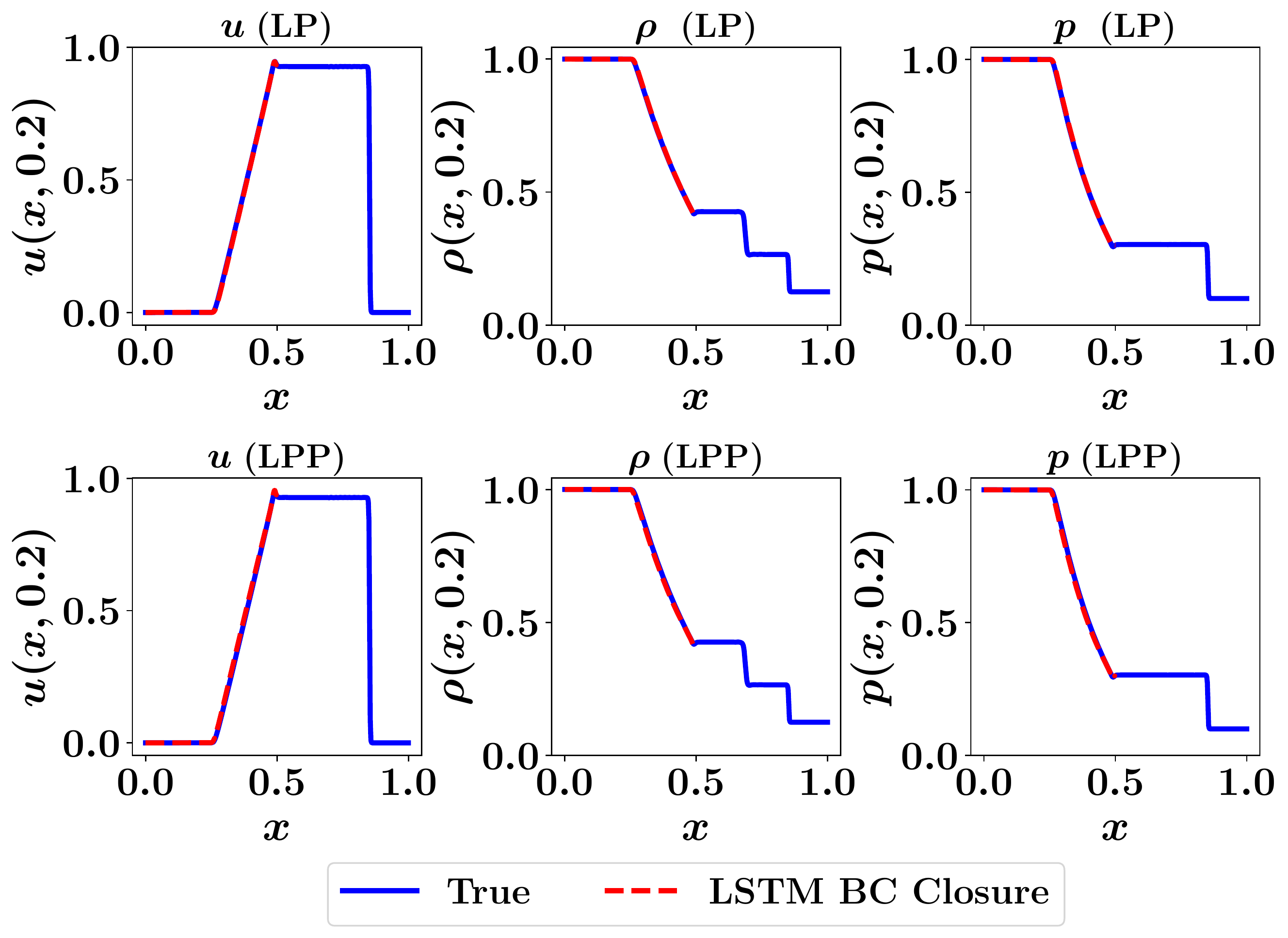}
\caption{\textcolor{rev}{Results for LSTM boundary condition closure for the Sod's shock tube at time $t=0.2$ with LP (top) and LPP (below) implementations.}}
\label{fig:res3}
\end{figure}

First, we apply the interface learning approaches with the interface at $x_b=0.5$ and consider the left zone to demonstrate the uplift learning concept. Results at final time (i.e. $t=0.2$) with both the LP and LPP implementations are given in Fig.~\ref{fig:res3}, which shows the success of the upwind learning framework to infer valid boundary conditions at the interface from the dynamics and flow pattern of its left-sided neighbors.

\begin{figure}[htbp!]
\centering
\includegraphics[trim= 0 0 0 0, clip, width=0.49\textwidth]{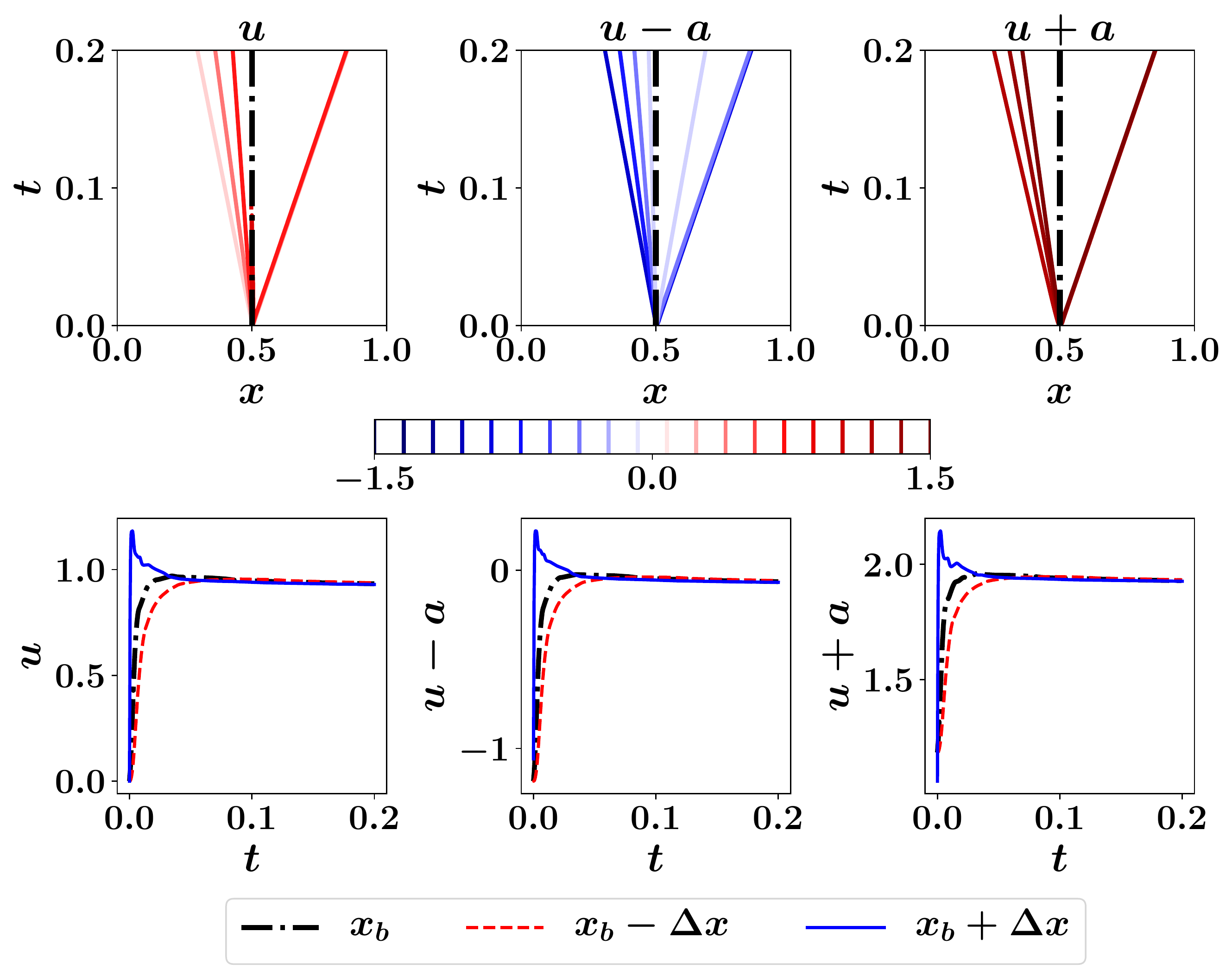}
\caption{\textcolor{rev2}{Contour plots of characteristics directions (top) as well as their time variation at $x_b=0.5$ (bottom) for the Sod's shock tube problem.}}
\label{fig:char}
\end{figure}

\textcolor{rev2}{In order to understand the performance of the upwind learning, we recall that the directions of characteristics (i.e., the curves $dx/dt=\lambda$ along which the Riemann invariants are constant) are defined as $\lambda_1=u$, $\lambda_2=u-a$, and $u+a$, where $a$ is the local speed of sound given as $a=\sqrt{\gamma (p/\rho)}$. We plot the space-time contour plots of $u$, $u-a$, and $u+a$ in Fig.~\ref{fig:char}, along with the line plot of the characteristics directions at the interface location. We observe that for $x=0.5$, $u$ and $u+a$ are always non-negative, while $u-a$ is initially negative (since the gas is initially at rest), but quickly approaches zero. In other words, the majority of the information at $x=0.5$ flows from left to right, and hence the success of the upwind learning scheme. Moreover, we remark that the interface conditions are almost constant with time, except for the initial transition period following the breakdown of the diaphragm, which minimizes the computational burden on the LSTM model.}

\begin{figure}[htbp!]
\centering
\includegraphics[trim= 0 0 0 0, clip, width=0.49\textwidth]{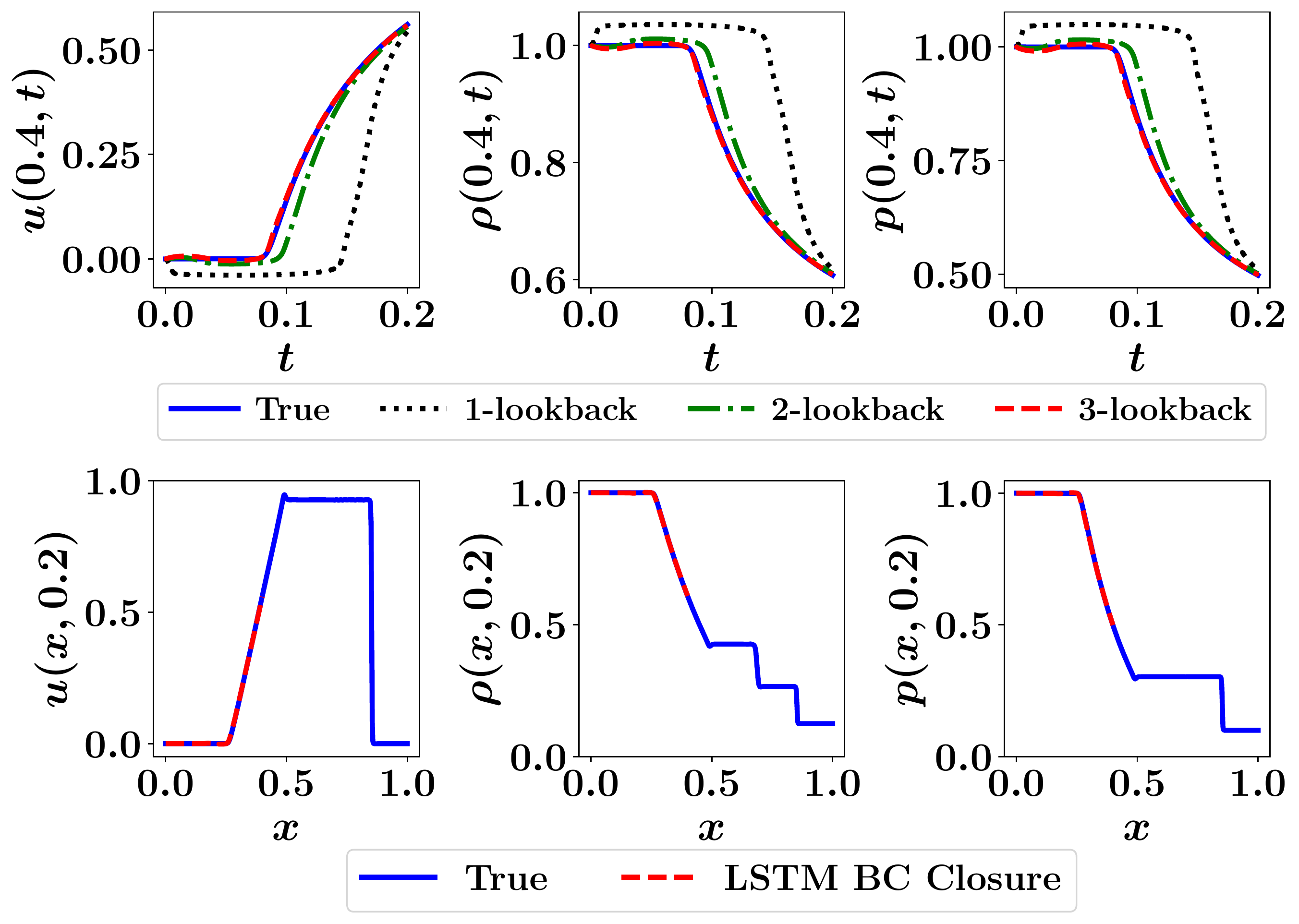}
\caption{\textcolor{rev2}{Time evolution of velocity, density, and pressure at $x_b=0.4$ with different lookback lengths (top), and their spatial distribution at final time with the LP approach using a lookback of 3 steps (bottom).}}
\label{fig:inter}
\end{figure}

\textcolor{rev2}{It can be seen from Fig.~\ref{fig:char} that moving the interface to the left will result in $u-a$ being significantly negative. In other words, some of the information should be inferred from the truncated region and the ML model has to account for this contribution. Indeed, we find that the same architecture with a single step lookback suffers in learning the interface conditions at $x_b=0.4$. However, augmenting the upwind learning framework with an increased time history of the modeled quantities (i.e., using a lookback of 3 steps) significantly improves the predictive capability as seen in Fig.~\ref{fig:inter}. That is enforcing a deeper time dependence facilitates learning the missing information due to domain truncation. Fig.~\ref{fig:inter} also depicts the spatial distribution of the velocity $u$, the density $\rho$ and the pressure $p$ at final time using the LP approach, augmented with a history of 3 time steps.} 

\textcolor{rev}{
\emph{Example 4: Fluid structure interaction in network flows}. To demonstrate the feasibility of the upwind interface learning in network domains, we construct a bifurcating flow in elastic tubes. This system is ubiquitous in cardiovascular system modeling \cite{stergiopulos1992computer,kissas2020machine} and open channel networks \cite{garcia1992mccormack,abhyankar2020petsc}, and it is often represented by the Saint-Venant equations. For such problems, a boundary closure issue appears at the bifurcation points and constitutive relations have to be imposed at these locations. Most often, this yields a system of nonlinear equations, which has to be solved with iterative schemes (e.g., Newton-Raphson method). \textcolor{rev2}{This incurs an additional computational cost to solve the system of nonlinear equations. It also requires careful selection of the numerical scheme to solve this system in order to guarantee its convergence after a few iterations.} Therefore, adopting the interface learning technique for these \textcolor{rev2}{systems} has the potential to address the bifurcation points treatment. Moreover, when the network grows largely, a large number of segments have to be considered simultaneously. However, with interface learning, the insignificant downstream segments can be truncated and their effects are modeled by the adopted ML architecture.}

\begin{figure}[htbp!]
\centering
\includegraphics[trim= 0 0 20 0, clip, width=0.49\textwidth]{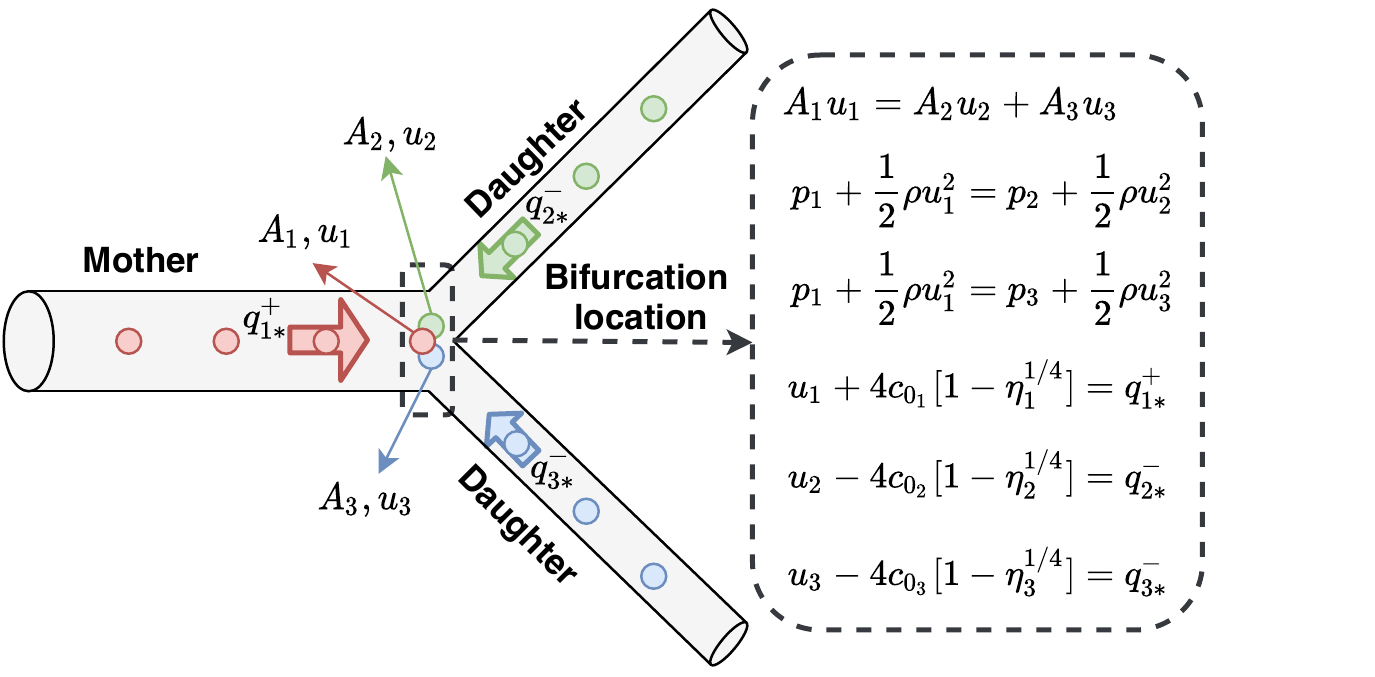}
\caption{\textcolor{rev2}{Problem setup for the bifurcating flow example. Subscripts $1$, $2$, and $3$ refer to the mother, first daughter, and second daughter, respectively, while the subscript $*$ denotes the nearest point to the bifurcation location (i.e., the second grid point for the daughter branches and second-to-last grid point for the mother segment).}}
\label{fig:bif}
\end{figure}

In this example, we consider a 3-segment network with a single bifurcation point with one mother (upstream) segment and two daughter (downstream) segments, governed by the following pulsed flow equations
\begin{align}
    \dfrac{\partial A}{\partial t} + \dfrac{\partial (u A) }{\partial x} &=0, \\
    \dfrac{\partial u}{\partial t} + \alpha u \dfrac{\partial u}{\partial x} &= -\dfrac{1}{\rho} \dfrac{\partial p}{\partial x} + \nu \dfrac{\partial^2 u}{\partial x^2} - \beta \pi \nu \dfrac{u}{A},
\end{align}
where $\rho$ and $\nu$ are the density and kinematic viscosity of the fluid, and $\alpha$ and $\beta$ are parameters depending on the assumed \emph{radial} velocity profile \cite{san2012improved}. Here, $\rho = 10^3$ kg/m$^3$, $\nu = 10^{-6}$ m$^2$/s, $\alpha = 1$, and $\beta = 8$. \textcolor{rev2}{A linear theory of elasticity can be used to relate the pressure and cross sectional area via $p=p_0 + 2\rho c_{0}^{2} (1-\sqrt{\eta})$, where $c_{0}$ is the wave propagation speed prescribed by the Moens-Korteweg equation, $c_{0}=\sqrt{E h /(2\rho R_0 (1- v^2))}$, and $\eta = A_0/A$ with $A_0=\pi R_{0}^{2}$ being the nominal reference value of cross sectional area when the pressure is $p_0$.} Here, $E$, $v$, and $h$ refer to the elastic modulus, Poisson ratio and thickness of the tube. \textcolor{rev2}{A schematic diagram of the problem setup we are solving is depicted in Fig.~\ref{fig:bif}, where the length ($L$) and nominal radius ($R_0$) of each segment are given in Table~\ref{table:network}. At bifurcation, six quantities become unknown (i.e., the ending cross sectional area and velocity of mother segment and the starting cross sectional area and velocity for each daughter branch. Continuity and total pressure equivalence constitute three equations, and the remaining three equations might come from characteristics (i.e., Riemann invariants). For pulsating flow equations, the Riemann invariants are approximated as follows \cite{san2012improved},
\begin{equation}
    q^{\pm} = u \pm 4c_0[1-\eta^{1/4}],
\end{equation}
where the plus-minus sign defines the direction of the characteristics. Within the mother segment, the information moves from the interior points to the bifurcation point through the right-travelling wave (i.e, with positive sign). Similarly, within each daughter branch, information flows from interior points (i.e, right-hand side points) to the bifurcation point via the backward-travelling wave (i.e., with negative sign). This forms a total of six nonlinear equations that can be solved using Newton-type methods as discussed above.}


\begin{figure}[htbp!]
\centering
\includegraphics[trim= 0 0 0 0, clip, width=0.49\textwidth]{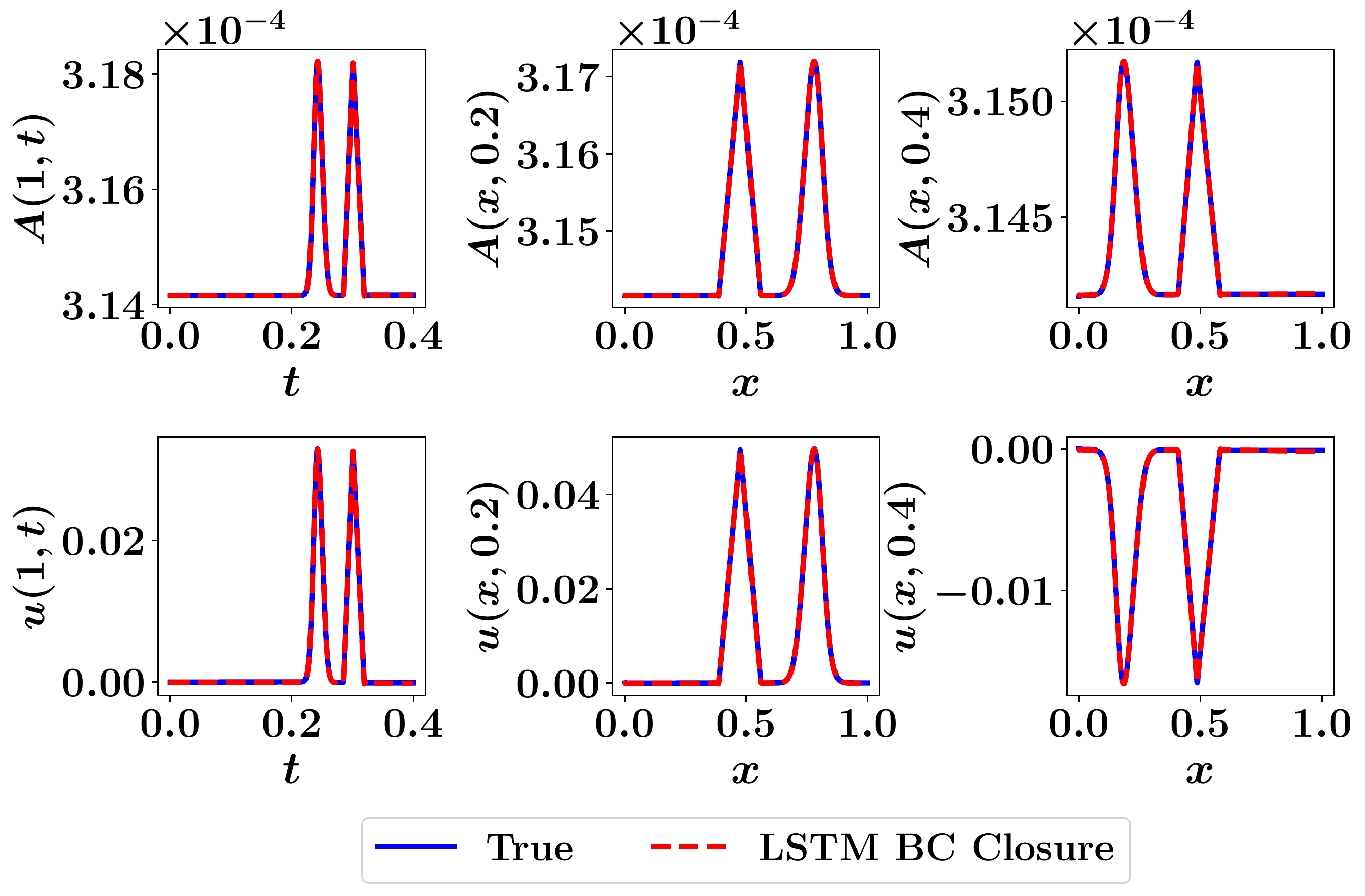}
\caption{\textcolor{rev}{Results for cross sectional area (top) and axial velocity (bottom) in the mother branch for the flow through a network of branched elastic tubes using the upwind learning compared to reference values obtained by solving the whole network and the bifurcation point. Subfigures in the middle and right columns illustrate the wave structures at $t=0.2$ s (before reaching to the bifurcation point) and $t=0.4$ s (after reflected from the bifurcation point), respectively.}}
\label{fig:res4}
\end{figure}

Initial conditions read as $A(x,0) = A_0$ and $u(x,0)=0$ for all segments. A composite Gaussian and triangular input wave signal is given as a boundary condition to the mother segment as follows:
\begin{align}
    u(0,t) &= 0.05 \max\bigg(e^{-k_1(t-k_2)^2},1-\bigg|\frac{t-k_2-3.5 k_3}{k_3}\bigg|,0\bigg), \\
    A(0,t) &= A_0 \bigg(1-\frac{u(0,t)}{4 c_0}\bigg)^{-4},
\end{align}
where $k_1 = 10^4$ s$^{-2}$, $k_2=0.05$ s, and $k_3=k_2/3$ \textcolor{rev2}{and reflecting outflow boundary conditions are imposed at the end of daughter segments.} \textcolor{rev2}{The} second order Lax-Wendroff scheme is followed to collect equispaced $40,000$ time snapshots for a maximum time of $0.4$ collected with a spatial step size of $10^{-4}$. For the upwind learning, we only solve for the mother branch and use an LSTM at the bifurcation point to model the effects of the daughter branches. The obtained results for the cross sectional area and axial velocity are given in Fig.~\ref{fig:res4} for the LP implementation in order to demonstrate the feasibility of the proposed approach. Once the mother segment bifurcates to the daughters of its half radius, the wave with one-third of its speed will be reflected from \textcolor{rev2}{the} bifurcation point. As shown in Fig.~\ref{fig:res4}, the reflected pulse has been modeled accurately by LSTM boundary conditions (notice the middle and right panels of the bottom row).

\begin{table}[htbp!]
\caption{\textcolor{rev}{Properties of the 3-segment branching network.}}
\centering
\begin{tabular}{p{0.09\textwidth}p{0.07\textwidth}p{0.07\textwidth}p{0.08\textwidth}p{0.07\textwidth}p{0.05\textwidth}}  
\hline
 Segment & $L$ (m) & $R_0$ (cm) & $E$ (MPa) &  $h$ (cm) & $v$  \\
\hline \smallskip \\
Mother & $1.0$   & $1.0$ & $0.4$   & $0.10$ & $0.5$ \\ 
Daughter & $1.0$   & $0.5$ & $0.4$   & $0.05$ & $0.5$ \\
Daughter & $1.0$   & $0.5$ & $0.4$   & $0.05$ & $0.5$ \\ 
\hline
\end{tabular}
\label{table:network}
\end{table}

\emph{Conclusions} --- 
In this work, we demonstrate the potential of machine learning tools to advance and facilitate CFD simulations of multiscale, multicomponent systems. In particular, we show the capability of memory embedding to learn the dynamics at the interface between different zones. This is especially beneficial where the domain contains zones with strong dynamics and components with complex configuration that might dictate a very fine mesh resolutions and time stepping. The proposed approach enables us to focus our efforts onto the domain portion of interest, while satisfying physically consistent interface conditions. It can serve as a non-iterative domain decomposition method. Toward model fusion technologies, such an interface learning methodology might also hold significant promise for the development of blending criteria in hybrid RANS/LES models. A proof-of-concept is first demonstrated using the 1D viscous Burgers equation over a two-zone domain with different physical parameters. An LSTM is used to bypass the micro-solver corresponding the stiff region and provide valid interface boundary conditions to enable the macro-solver to run independently. \textcolor{rev}{To consider hyperbolicity and wave structures, we furthermore propose the concept of \emph{upwind learning} towards a physics-informed domain decomposition, with illustrations using a shock tube problem in gas dynamics and a fluid-structure interaction application in network flows.} We illustrate the success and robustness of the proposed methodology using different learning configurations. Both LP and LLP models are the key concepts introduced in this letter, especially for designing intelligent boundary closure schemes, which may bear huge potential in many scientific disciplines. \textcolor{rev2}{Moreover, we demonstrate that the performance of interface learning architectures can be significantly improved by increasing the length of lookbacks (i.e., enforcing deeper time history).}  Finally, we emphasize that a similar interface closure technique can be adopted in high performance computing environments, to minimize the communication cost and delay between different asynchronous processors, a topic we would like to pursue further in the future.  

\textcolor{rev2}{We appreciate the valuable comments raised by the anonymous reviewers. Their suggestions have significantly helped us to improve the quality and clarity of the paper.} This material is based upon work supported by the U.S. Department of Energy, Office of Science, Office of Advanced Scientific Computing Research under Award Number DE-SC0019290. O.S. gratefully acknowledges their support.

The data that support the findings of this study are available from the corresponding author upon request. \textcolor{rev}{Python implementation to reproduce the results presented in this letter can be accessed from the Github repository (\href{https://github.com/Shady-Ahmed/UpwindLearning}{https://github.com/Shady-Ahmed/UpwindLearning}).}





\bibliography{references}
\end{document}